\documentclass[aps,prl,twocolumn,showpacs]{revtex4-1}
\usepackage{graphicx}
\usepackage{bm}
\usepackage{wasysym}
\usepackage{setspace}

\begin{document}

\title{Non-magnetic Stern-Gerlach Experiment from Electron Diffraction}
\author{Chyh-Hong Chern}
\email{chchern@ntu.edu.tw} 
\author{Cheng-Ju Lin}
\affiliation{Department of Physics and Center for Theoretical Sciences, National Taiwan University, Taipei 10617, Taiwan}
\date{\today}
\begin{abstract}
Using the wave nature of the electrons, we demonstrate that a transverse spin current can be generated simply by the diffraction through a single slit in the spin-orbital coupling system of the two-dimensional electron gas. The diffracted electron picks up the transverse momentum.  The up spin electron goes one way and the down spin electron goes the other, producing the coherent spin current.  In the system of spin-orbital coupling $\sim10^{-13}$ eV$\cdot$m, the \emph{out-of-plane} component of the spin of the electron can be generated up to 0.42 $\hbar$.  Based on this effect, a novel device of grating to distill spin is designed.  Two first diffraction peaks of electron carry different spins, duplicating the non-magnetic version of Stern-Gerlach experiment.  The direction of the spin current can be controlled by the gate voltage with low energy cost. 
 \end{abstract}

\maketitle


One of the core principles in quantum mechanics is the quantum superposition principle stemming from the wave nature of matter.  Especially, when Feynman taught his path integral, one of his favorite experiments to illustrate this powerful principle was the double-slit interference of electrons$^1$.  The interference pattern of electrons through the double slits is determined by the square of the absolute value of the quantum amplitudes $\psi_1+\psi_2$, where $\psi_k$ is the amplitude from the source to the screen passing through the $k^{\textrm{th}}$ slit.  In his note, electron is treated as a spinless particle, and the quantum amplitude $\psi_k$ is the phasor, originated from the optics, the complex number of unit absolute value.  His approach is correct in the sense that electrons in the free space are spin unpolarized and it is not necessary to take electron spin into consideration.  Therefore, the interference pattern of electrons is identical to that for photons.  However, the situation is quite different when considering the electrons propagating in the materials of non-vanishing spin-orbital (SO) interaction, where the angle between the spin orientation and the propagation direction is fixed.  Different types of SO couplings yield different correlations between the spin orientation and the propagation direction.  The spin, thus, becomes an important degree of freedom, and up spins and down spins could have different responses to the diffraction.

Recently, the systems of spin-orbital coupling have been attracting great attention to people due to their great potential in spintronics applications as well as the realization of the quantum computing$^{2-8}$.  The major breakthrough will critically reply on the technology to manipulate the electron spin and/or spin current.  The problems are two-fold.  One is to generate spin current, and the other is to control the direction of spin in the transportation.  In particular, the manipulations using the non-magnetic stimuli, such as electric field$^9$ or electric current, are considered most practical in the device applications.  Distilling up spin and down spin has been one of the most important challenges in physics.  In this Letter, we illustrate a new method of generating the spin current by the electron diffraction from a single slit in the 2-dimensional electron gas (2DEG) system.  The electrons of up-spin component and down-spin component go in the opposite transverse directions resulting in a significant spin-splitting effect.  This controllable and remarkable effect leads to an efficient method to separate up spin and down spin electrons \emph{without} magnetic field. Inspired by the grating for photons, we will show that a grating-like structure makes not only the spin-splitting effect but also the spin wave packets become \emph{solitonic}, mimic the Stern-Gerlach experiment \emph{without} magnetic field.  Furthermore, the direction of the spin current can be controlled by changing the chemical potential using the gate voltage in the range of $m$eV.

Let us consider a 2DEG in a semiconductor heterostructure connected with leads.  In the presence of SO coupling, the system can be described by the following two-dimensional Hamiltonian
\begin{eqnarray}
H=\frac{p^2}{2m^*}-\mu+\alpha(\hat{\sigma}_x p_y-\hat{\sigma}_y p_x)+\beta(\hat{\sigma}_xp_x-\hat{\sigma}_y p_y), \label{eq:hamiltonian}
\end{eqnarray}
where $m^*$ is the effective mass of electron, $\mu$ is the chemical potential, and $\hat{\sigma}_k$ are the Pauli spin matrices.  The spin-orbital coupling of the 2DEG in the semiconductor heterostructure is described by the $\alpha$ and $\beta$ terms in Eq.~(\ref{eq:hamiltonian}), called the Rashba and the Dresselhaus couplings respectively. The energy bands of Eq.~(\ref{eq:hamiltonian}) are given by $E^\pm_p=p^2/2m-\mu\pm\Delta_p$, where $\Delta_p=\sqrt{(\alpha^2+\beta^2)p^2+4\alpha\beta p_xp_y}$ and $\pm$ are the band indices labeling the upper (+) and the lower (-) bands.  The electron spin described by Eq.~(\ref{eq:hamiltonian}) lies in the $xy$ plane and does not have the $z$ component.  The spin orientation is correlated by the propagation direction by $\phi_+ = \tan^{-1} (-\frac{\alpha p_x + \beta p_y}{\beta p_x + \alpha p_y})$ for the upper band and $\phi_-=\phi_+ +\pi$ for the lower band, where $\phi_\pm$ is the angle measured from the $x-$axis.  Note that the band structures of a pure Rashba system and a pure Dresselhaus system are the same.

The diffraction pattern is the superposition of the quantum waves from the slit.  When the electron reaches the slit, each point in the slit is considered as a point source of a new spherical wave known as the Huygens' principle.  The wave amplitude at a certain point on the screen is the quantum superposition of all the spherical wave emitted from the slit.   The quantum amplitude of the electron from $(x',y')$ at $t=0$ to $(x,y)$ at time $t$ denoted by $\langle x,y,t|x',y',0\rangle$ is called the kernel which is a $2\times 2$ matrix in our system.  Suppose the slit locates at $x=0$, and the screen is placed at $x=L$ away, the wavefunction on the screen is given by
\begin{eqnarray}
\psi(L,y,t) = \int_{-\frac{d}{2}}^{\frac{d}{2}} dy' \langle L,y,t|0,y',0\rangle \phi(0,y',0), \label{eq:diffraction}
\end{eqnarray}
where $d$ is the aperture size of the slit, and the center of the slit aligns at $y'=0$.  The $\phi(x',y',0)$ in Eq.~(\ref{eq:diffraction}) is the normalized initial wavefunction of the electron.  If the slit is small and reasonably thick, only the wave of momentum in the $x-$direction can pass through it.  Considering propagating in the $+x$-direction, $\phi(x',y',0)$, a two-component spinor denoted by $(\phi_1, \phi_2)$, is taken to be uniform in $y'$ in the slit as $\frac{1}{\sqrt{2d}}(i, -1)^{\textrm{T}}e^{\frac{i}{\hbar}p_F x'}$ in the lower band of the Rashba system and $\frac{1}{\sqrt{2d}}(1, 1)^{\textrm{T}}e^{\frac{i}{\hbar}p_F x'}$ in the upper band of the Dresselhaus system, where $p_F$ is the momentum at the Fermi energy.   In the following, $\psi(L,y,t)$ is computed for a pure Rashba system and for a pure Dresselhaus system separately.

The diffraction pattern on the screen is determined by $|\psi(y)|^2\equiv|\psi_1(y)|^2+|\psi_2(y)|^2$, where $\psi_k$ is the $k^{\textmd{th}}$ component of the spinor $\psi(y)$.  The results of $|\psi(\bar{y})|^2$ for the Rashba and the Dresselhaus systems are given in Fig.~(1a), where  the dimensionless coordinate $\bar{y}=y/L$ is used.  The $|\psi(y)|^2$ is computed perturbatively to the fifth order in $\bar{\alpha}=\frac{m^*\alpha L}{\hbar}$ (or $\bar{\beta}$).  In the real systems$^{10, 11}$, the slit acts like a momentum selector.   As long as the thickness of the slit is greater than $\lambda_F$, where $\lambda_F$ is the Fermi wavelength, only the electron of the $\pm x$ momentum can pass through.  In Fig.~(1a), our results of $|\psi(\bar{y})|^2$ has no difference from the system without SO coupling:  The first dark fringe occurs at $\frac{d}{\lambda_F}\frac{\bar{y}}{\sqrt{1+\bar{y}^2}}=1$, so is the relative brightness between the higher-order bright fringes and the central peak.

\begin{figure}[htb]
\includegraphics[width=0.45\textwidth]{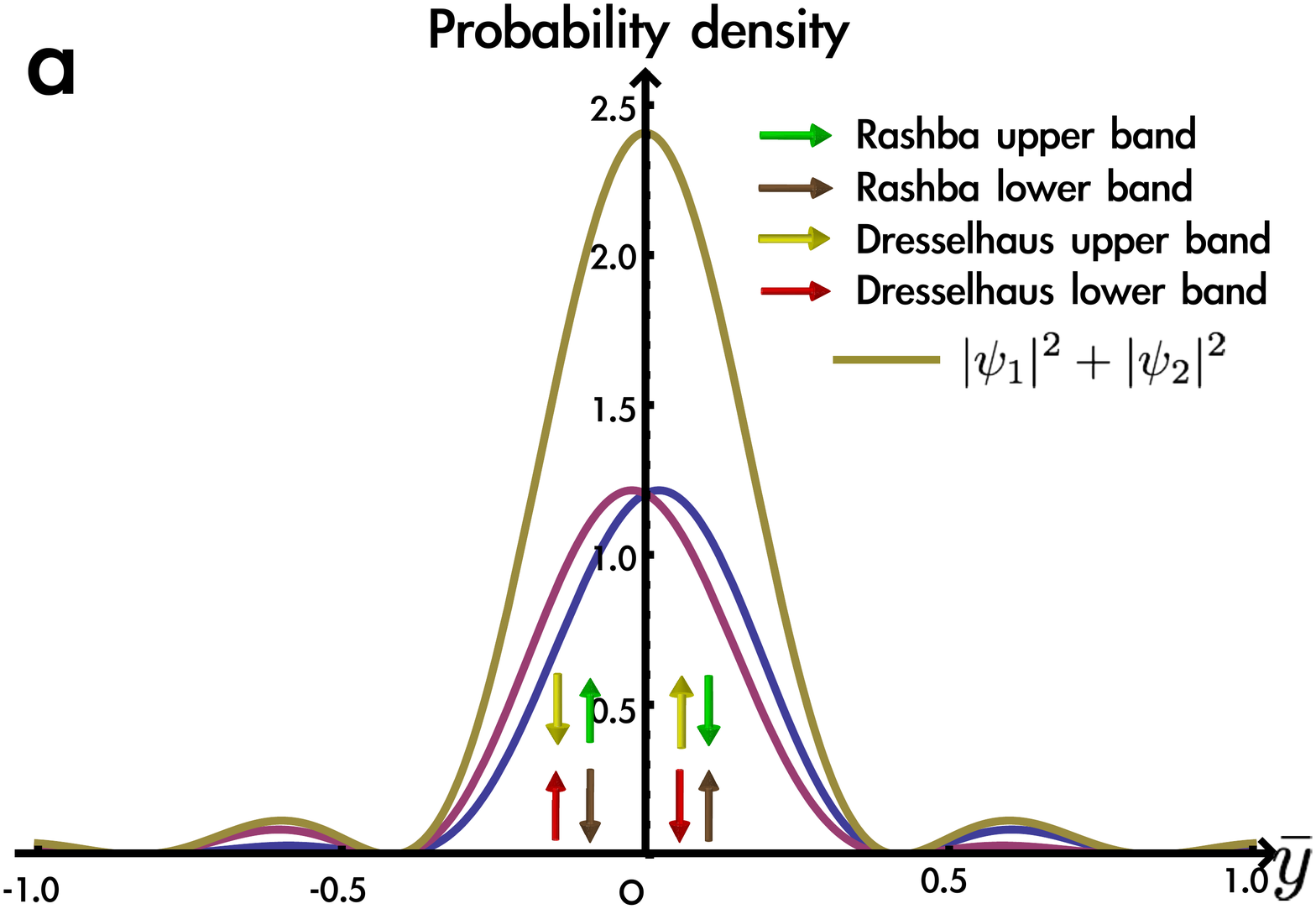}
\includegraphics[width=0.45\textwidth]{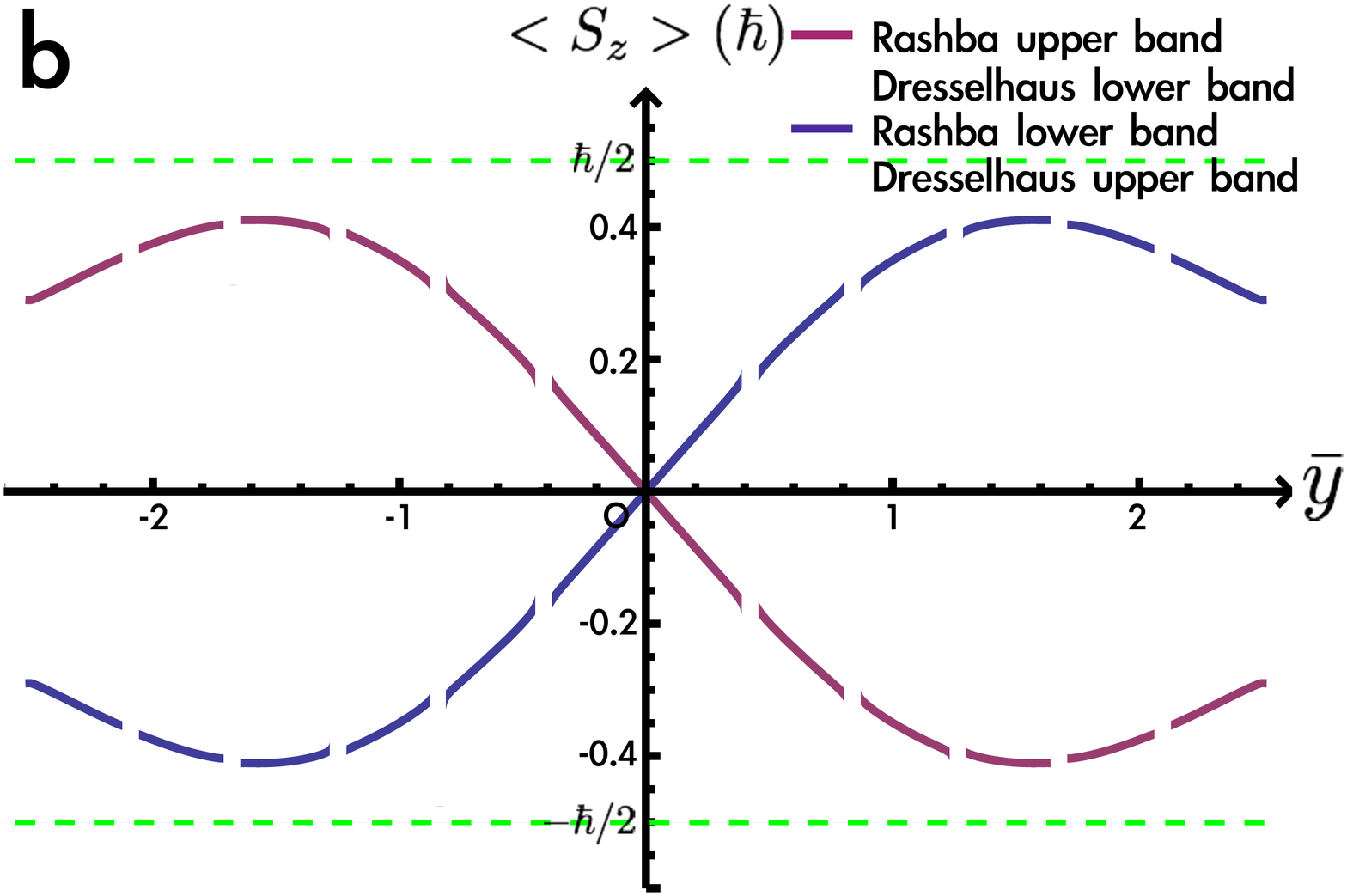}
\caption{(color online) (a) The probability density of $|\psi(\bar{y})|^2$ and $|\psi_k(\bar{y})|^2$ of the diffraction through a single slit for the upper and lower bands of the Rashba and Dresselhaus systems.  This plot should be deciphered as the following: $|\psi(\bar{y})|^2$ are the same for all cases, and spin distribution $|\psi_k(\bar{y})|^2$ are different by cases.  For example, the spin of the upper band of the Rashba system is marked in green.  The probability density marked in red denotes $|\psi_1(\bar{y})|^2$ and the one marked in blue labels $|\psi_2(\bar{y})|^2$.  (b)  The spin distribution of the out-of-plane component $\langle S^z(\bar{y})\rangle$. $\langle S^z\rangle$ increases with the diffraction angle.  At $\bar{y}=0$, the fictitious magnetic field is parallel to the spin orientation of the initial wavefunction, so it remains in the plane.  At $\bar{y}\neq0$, they are not parallel, leading to  the spin precession to produce $\langle S^z\rangle$.  The angle between them increases with the diffraction angle, so $\langle S^z\rangle$ increases with $\bar{y}$.  Furthermore, spin propagating in different $y$-direction precesses in opposite direction, leading to the remarkable spin-splitting effect.  In this plot, $\langle S^z(\bar{y})\rangle$ in all cases are calculated.}\label{Fig:system1}
\end{figure}

The major difference lies in the spin distribution.  The $|\psi_1(\bar{y})|^2$ and the $|\psi_2(\bar{y})|^2$ have the asymmetric diffraction patterns shown in Fig.~(1a).  The positions of the dark fringes of the $|\psi_1(\bar{y})|^2$ and the $|\psi_2(\bar{y})|^2$ are the same as the $|\psi(\bar{y})|^2$, since it is the property of the phase difference regardless the spinor part of the wavefunction.  We note that the electron spin in these systems is parallel to the $xy$ plane, so $|\phi_1(\bar{y})|^2=|\phi_2(\bar{y})|^2$.  That $|\psi_1(\bar{y})|^2\neq |\psi_2(\bar{y})|^2$ indicates that up spins and down spins favor different propagation directions.  Taking the electrons in the lower band of the Rashba system as the example, the up spin favors the positive $y$-direction, and the down spin favors the minus $y$-direction.  In other words, after the diffraction, the electron spin picks up the \emph{out-of-plane} component and spin splits in the real space.  

Our results imply that spin current is generated by the diffraction.  To see this more clearly, we plot $\langle S^z(\bar{y})\rangle$ in the Fig.~(1b).  Taking the lower band of the Rashba system as the example, the electron observed in the positive (minus) $y$-direction is $\langle S^z\rangle > 0 (< 0)$.  Therefore, after diffraction, the up (down) spin picks up a velocity in the positive (negative) $y$ direction.  If we define the spin current to be $I^i_j\equiv\langle S^i\rangle v_j$, this result implies $I^z_y\neq 0$.  In addition, because $|\psi(y)|^2$ is the even function in $y$, the electric current $I_y$ is zero.  Noting that a pure spin-up electron is $\langle S^z\rangle = \frac{1}{2}\hbar$, our results show the maximum spin polarization goes up to 84\%.  The functional form of $\langle S^z(\bar{y})\rangle$ depends on $L$ and the strength of the SO coupling.  It is also different in different bands and systems.  Surprisingly, it does not depend on the wavelength of the electron nor the aperture size of the slit.  Therefore, there is a transition of $\langle S^z(\bar{y})\rangle$ if the chemical potential is tuned between upper and lower bands.

The direction of $I^z_y$ can be changed by tuning the chemical potential.  In both Rashba and Dresselhaus systems, the spin distribution changes sign between two different bands, indicating the reversal of the direction of the spin current.  One can change the chemical potential to tune the Fermi level between the upper and the lower bands that touch at $(p_x,p_y)=(0,0)$.  If one tunes the chemical potential above the band touching point, the electrons  of two momenta will pass the slit.  One is from the upper band, and the other is from the lower band.  Since the one from the lower band has shorter wavelength, the effect of diffraction is smaller than the one from the upper band, so $I^z_y > 0$.  Similarly, if the chemical potential is tuned below the band touching point, the electron of the larger wavelength has larger effect in the lower band, so $I^z_y<0$.  Therefore, the value of the chemical potential determines the direction of the spin current, that can be controlled by applying a back gate voltage in experiments.

The novel spin-splitting effect can be understood as the following.  In the SO coupling system, the direction of the spin orientation is locked with the propagation direction.  This property can be effectively thought of the presence of a fictitious magnetic field accompanying with the electron, which tends to lock the electron spin.  In the initial wavefunction, the spin orientation is pointing along the $y$ or $x$ directions depending on whether it is the Rashba system or the Dresselhaus system.  After the electron is diffracted by the slit, it feels the fictitious magnetic field.  The direction of the fictitious magnetic field is different from the spin orientation of the initial wavefunction in general.  Therefore, the spin precesses to result in the inhomogeneous spin distribution.  For higher diffraction angle, the angle between the spin orientation and the fictitious magnetic field is usually larger, and it produces faster rate of precession.  Therefore, $\langle S^z(\bar{y})\rangle$ increases with $\bar{y}$ as shown in Fig.~(1b).  It also explains the different spin distributions in Rashba and the Dresselhaus systems.

\begin{figure}[htb]
\includegraphics[width=0.45\textwidth]{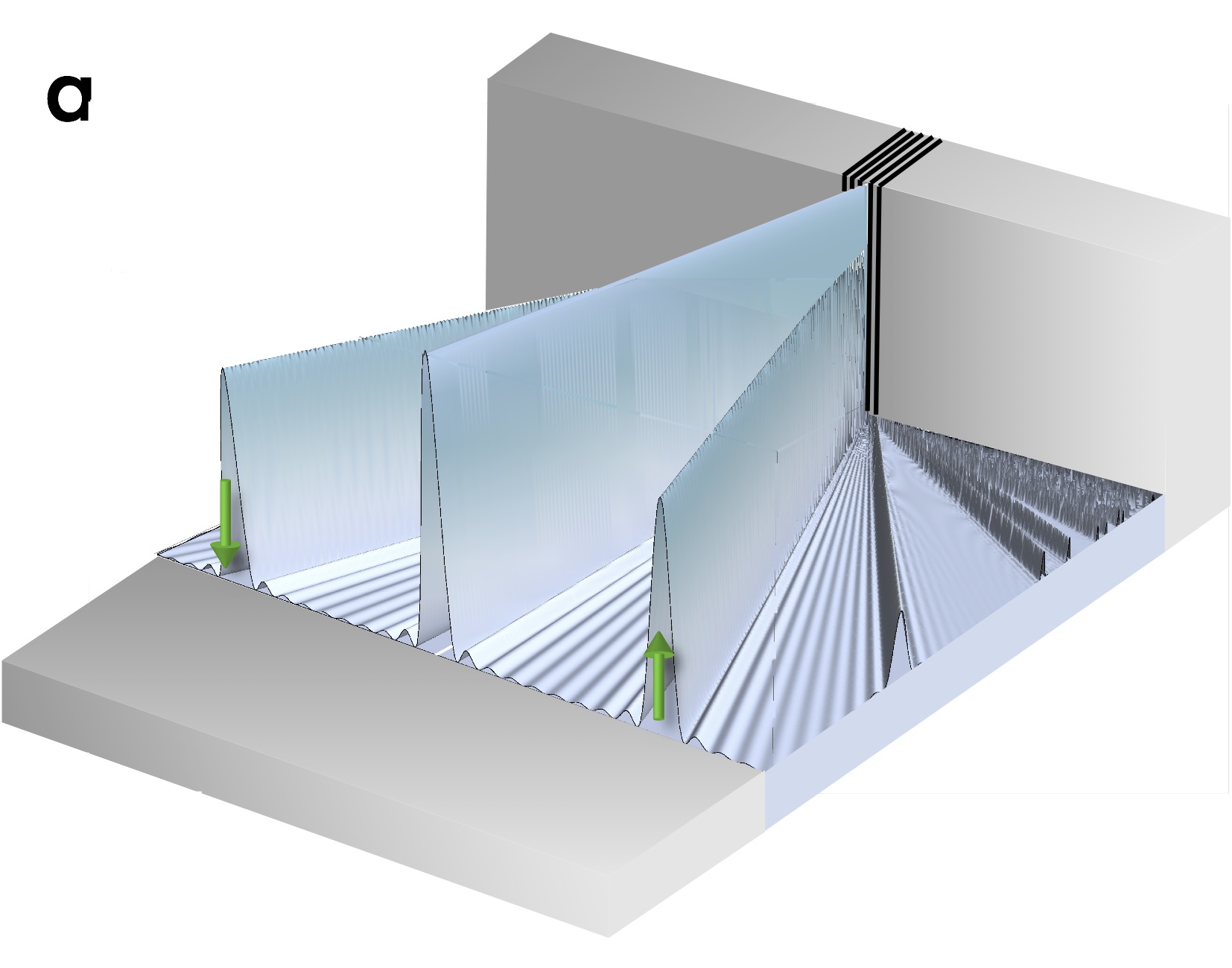}
\includegraphics[width=0.45\textwidth]{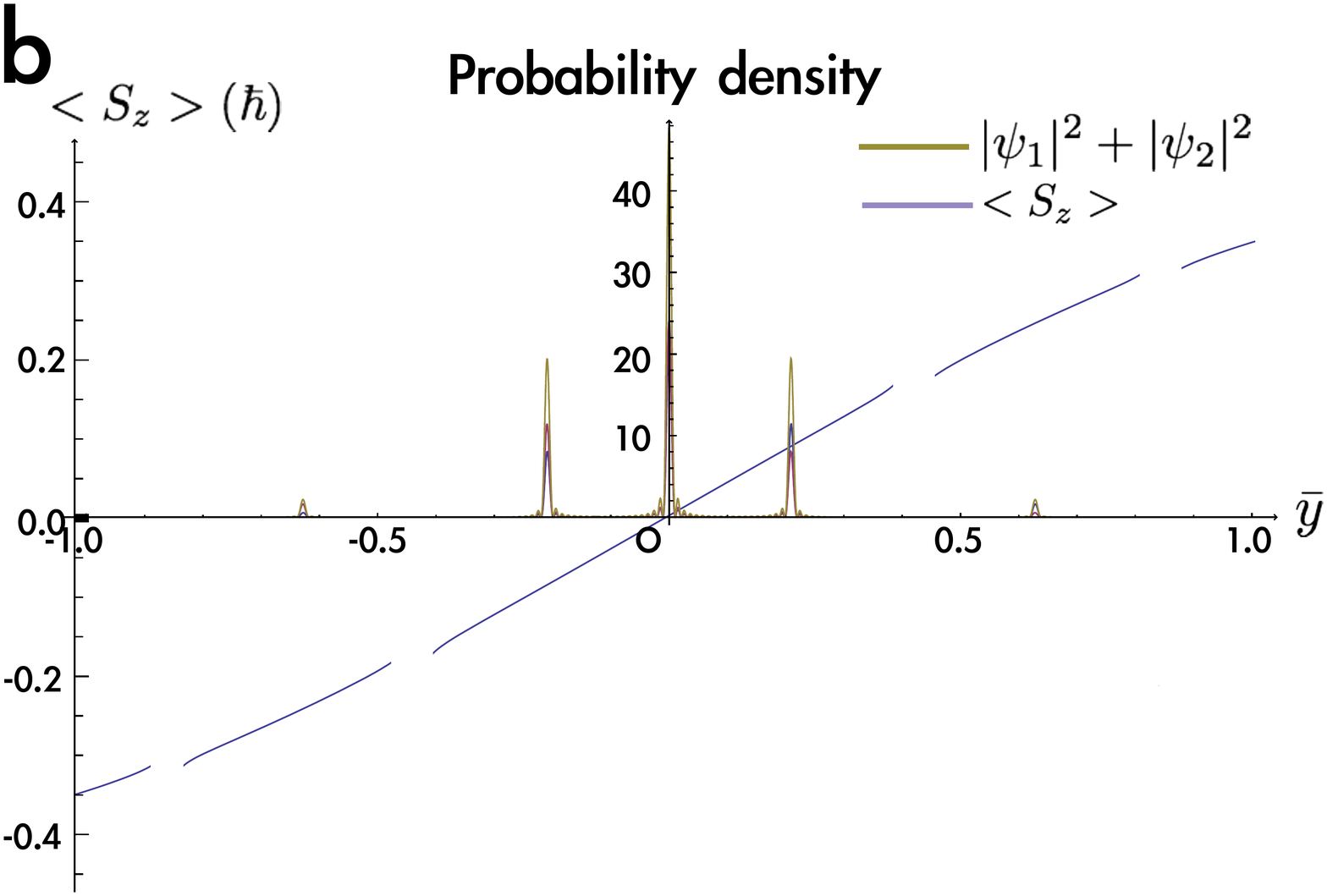}
\caption{(color online) (a) The overview of the solitonic wavepackets produced by the grating.  (b) The probability density of $|\psi(\bar{y})|^2$ and $|\psi_k(\bar{y})|^2$ of the diffraction through a grating with number of slits $N=20$.  The aperture size of each slit $\bar{d}$ is given in the text.  The distance between the centers of the nearest neighbor slits is also $\bar{d}$.  Similar to the grating diffraction in optics, the electron wavepackets are sharp and well separated.  The blue curve is the $\langle S^z(\bar{y})\rangle$ measured by the left axis, which is same as the one for the single slit, because it is independent of the number of slits and the wavelength of the electrons.  Using these properties, one can tune the chemical potential so that the first diffraction peak locates near the maximum of $\langle S^z\rangle$ to achieve maximum spin polarization and further design a grating so that $\langle S^z\rangle_{\textrm{max}}\sim\frac{1}{2}\hbar$.}\label{Fig:system2}
\end{figure}

The closest experimental setup to our proposal in the current semiconductor devices may be the quantum point contact.  The diffraction fringes of the coherent 2DEG by the quantum point contact has been imaged using the Cryogenic Scanning Probe Microscopes (SPM)$^{10-13}$.  To demonstrate the robustness of our effect, we put some numbers accessible in the experimental range.  Suppose the chemical potential is tuned so that the Fermi wave number is $0.02$ $\mathring{\textmd{A}}^{-1}$, the width of the slit is 79 nm, and the $L=7.9$ $\mu$m, spin-splitting between the highest peaks of up spin and down spin can be 2.4 $\mu$m if the Rashba (or Dresselhaus) parameter $\alpha\hbar$ (or $\beta\hbar$) is $\sim 10^{-13}$ eV$\cdot$m provided that the effective mass of the electron $m^*$ is $0.05$ $m_{e}$, where $m_e$ is the bare mass of the electron.  In these parameters, the first dark fringe occurs roughly at $0.4L=3.1$ $\mu$m away from the slit center in the $y$ direction, and the first peak of the $\langle S^z\rangle$ occurs roughly at $0.15L=1.2$ $\mu$m.  We note that the Rashba parameter used in the estimation is not particularly large.  It can be as large as $4\times 10^{-11}$ eV$\cdot$m in the InAs.

In the way to realize the spintronics devices and the quantum computing, one of the most crucial steps is to separate up spin and down spin in the materials.  In the early twenty centuries, Stern and Gerlach separated them by using an inhomogeneous magnetic field in the free space.  Based on the current effect, a grating-like structure can do this job.  The system overview is shown in Fig.~(2a).  Fig.~(2b) shows the probability density of the electron diffraction through a grating of number of slits $N=20$ and the $\langle S^z(\bar{y})\rangle$ in the lower band of the Rashba system.  The probability density of an electron in Fig.~(2b) makes no difference from that of a photon.  It is easy to read $\langle S^z\rangle$ of the diffraction peaks.  The central peaks is $\langle S^z\rangle=0$, for example.  The up spin and down spin are differentiated by the first diffraction peaks.  The up (down) spin goes in the positive (negative) $y$-direction.  As mentioned earlier, $\langle S^z(\bar{y}) \rangle$ does not depend on the wavelength of the electron, so one can change the positions of the diffraction peak so that it locates at the maximum of $\langle S^z\rangle$ by tuning the chemical potential.  Although it is not shown here,  the $\langle S^x(\bar{y}) \rangle$ and $\langle S^z(\bar{y}) \rangle$ are odd functions of $\bar{y}$ but the $\langle S^y(\bar{y}) \rangle$ is the even function.  One can fine tune the chemical potential so that the first diffraction peaks locate at $\langle S^y(\bar{y}) \rangle=0$ where  in our result is close to where the maximum $\langle S^z \rangle$ is.  Then, in the screen, two spin wavepackets with anti-parallel spin orientations at the first diffraction peaks can be seen, realizing the \emph{non-magnetic} Stern-Gerlach experiment.

In summary, a generalization of Feynman's favorite experiment of the electron diffraction leads to a new method of generating spin current in the SO coupling system.  The transverse spin current occurs naturally when 2DEG is diffracted through slits.  The direction of the spin current can be controlled by the gate voltage, offering a convenient way to manipulate spins and enhancing its potential in applications as well.  In addition, a grating that separates spin is analyzed.  This profound effect may stimulate the device design that would hopefully leads us to see the spintronics devices or quantum computing in service in the near future.


We appreciate the stimulated discussions with Chih-Wei Chang, Chi-Te Liang, and Minn-Tsong Lin for various experimental aspects.  We express the highest gratitude to Cheng-Hsuan Chen for proofreading our manuscript before its submission.  This work is supported by NSC 97-2112-M-002-027-MY3 of Taiwan.
%

\end{document}